\documentclass[12pt, a4paper, twoside, titlepage]{article}
\usepackage[latin1]{inputenc}
\usepackage[T1]{fontenc} 
\usepackage{graphicx}
\usepackage[table,xcdraw]{xcolor}
\usepackage{lineno,hyperref}
\usepackage{amsmath}
\usepackage{amssymb}

\begin{document}
\title{A versatile trend test for the evaluation of tumor incidences in long-term carcinogenicity bioassays}
\author{Ludwig A. Hothorn $^1$, Atiar M. Rahman$^2$ and Frank Schaarschmidt $^1$,\\
1) Leibniz University Hannover, Faculty of Natural Sciences,\\ Herrenhauser Str. 2, D-30419 Hannover, Germany\\
2) FDA, Center for Drug Evaluation, Office of Biostatistics, Silver Spring, USA}

\maketitle

\begin{abstract}
For the evaluation of carcinogenicity bioassays a new trend test is proposed
which is based on a maximum of arithmetic, ordinal, and logarithmic regression scores as well as the Williams-type contrasts for either crude proportions or more appropriate poly3-estimates for the tumor-by-time relationships. This test provides an almost appropriate power for most shapes of dose-response relationships (including for possible downturn effect at high(er) dose(s)), common signs of significance (p-value, confidence limits) and the information on the probable shape. Related software is easily available within the CRAN-packages \verb|tukeytrend, MCPAN, multcomp|
\end{abstract}

\section{Introduction}
Aim of a long-term carcinogenicity bioassay is the evaluation of the tumorigenic effect of a compound when it is 
administered to rodents for most of their life span \cite{Elwell2002, Morton2002}. 
Tumor-site specific incidences $ n_{\text{developing a tumor}}/ n_{\text{at risk}}$ are of primary interest. This means the evaluation of a 2-by-k contingency table for the randomized design including a negative control group and commonly three dose groups (per species and sex) $[p_0, p_1,p_2, ...,p_k]$.\\
 Because longer living animals may develop tumors more likely than those dying earlier, this evaluation of crude tumor proportions could be distorted. Instead, their mortality-adjusted analysis is recommended \cite{Kodell2012}. Historically, the joint testing of age-adjusted tumor lethality for fatal tumors and age-adjusted tumor prevalence for incidental tumors is performed \cite{Peto1980}. This requires strictly the availability of valid cause-of-death information for each individual tumor in each individual animal. An alternative without cause-of-death information is the poly-k adjustment \cite{Bailer1988}  where the number at risk is a certain function of animal-specific survival times within each dose group. According to several guidelines \cite{FDA2001}, \cite{NTP2013} this is the primary principle to identify a dose-dependent increasing trend commonly by the Cochran\textendash Armitage test and its poly-3 modifications \cite{Kodell2012}. However, this trend test has a power problem (i.e. an unnecessary high false negative rate) for selected non-linear shapes, particularly plateau-like shapes. Already the title of the original pioneering paper \cite{Armitage1955}. stated correctly \textit{A test for linear linear trends...}. Near-to-linear shapes occur, but real data show a much broader variety of shapes, e.g. in a rat lifetime bioassay on romosozumab \cite{Chouinard2016} selected tumor incidences are:

\begin{table}[h]
\centering
\caption{2-by-k contingency tables for rat bioassay on romosozumab}
\tiny
\label{my-label}
\begin{tabular}{lllll|llll}
                                             & \multicolumn{4}{l}{\textbf{Males}}                                                                                                                 & \multicolumn{4}{l}{\textbf{Females}}                                                                                                               \\
Dose in mg/kg/wk                             & 0                                  & 3                                   & 10                                 & 50                                 & 0                                  & 3                                   & 10                                 & 50                                 \\
animals exam.                                & 60                                 & 60                                  & 60                                 & 54                                 & 60                                 & 60                                  & 60                                 & 60                                 \\ \hline
\textit{Adrenal cortical adenoma}            & \cellcolor[HTML]{C0C0C0}\textbf{0} & \cellcolor[HTML]{C0C0C0}\textbf{2}  & \cellcolor[HTML]{C0C0C0}\textbf{6} & \cellcolor[HTML]{C0C0C0}\textbf{2} & \cellcolor[HTML]{C0C0C0}\textbf{6} & \cellcolor[HTML]{C0C0C0}\textbf{3}  & \cellcolor[HTML]{C0C0C0}\textbf{1} & \cellcolor[HTML]{C0C0C0}\textbf{7} \\
\textit{Benign pheochromocytome in adrenals} & \cellcolor[HTML]{C0C0C0}\textbf{4} & \cellcolor[HTML]{C0C0C0}\textbf{3}  & \cellcolor[HTML]{C0C0C0}\textbf{6} & \cellcolor[HTML]{C0C0C0}\textbf{3} & \cellcolor[HTML]{C0C0C0}\textbf{2} & \cellcolor[HTML]{C0C0C0}\textbf{8}  & \cellcolor[HTML]{C0C0C0}\textbf{2} & \cellcolor[HTML]{C0C0C0}\textbf{0} \\
\textit{Renal lipoma in kidneys}             & \cellcolor[HTML]{C0C0C0}\textbf{0} & \cellcolor[HTML]{C0C0C0}\textbf{0}  & \cellcolor[HTML]{C0C0C0}\textbf{0} & \cellcolor[HTML]{C0C0C0}\textbf{1} & \cellcolor[HTML]{C0C0C0}\textbf{1} & \cellcolor[HTML]{C0C0C0}\textbf{0}  & \cellcolor[HTML]{C0C0C0}\textbf{0} & \cellcolor[HTML]{C0C0C0}\textbf{3} \\
\textit{C-cell adenoma in thyroid gland}     & \cellcolor[HTML]{C0C0C0}\textbf{8} & \cellcolor[HTML]{C0C0C0}\textbf{10} & \cellcolor[HTML]{C0C0C0}\textbf{4} & \cellcolor[HTML]{C0C0C0}\textbf{6} & \cellcolor[HTML]{C0C0C0}\textbf{5} & \cellcolor[HTML]{C0C0C0}\textbf{11} & \cellcolor[HTML]{C0C0C0}\textbf{8} & \cellcolor[HTML]{C0C0C0}\textbf{8} \\
\textit{Iselt cell adenoma in pancreas}      & \cellcolor[HTML]{C0C0C0}\textbf{1} & \cellcolor[HTML]{C0C0C0}\textbf{1}  & \cellcolor[HTML]{C0C0C0}\textbf{3} & \cellcolor[HTML]{C0C0C0}\textbf{4} & \cellcolor[HTML]{C0C0C0}\textbf{0} & \cellcolor[HTML]{C0C0C0}\textbf{0}  & \cellcolor[HTML]{C0C0C0}\textbf{3} & \cellcolor[HTML]{C0C0C0}\textbf{3} \\
\textit{Islet cell carcinoma in pancreas}    & \cellcolor[HTML]{C0C0C0}\textbf{0} & \cellcolor[HTML]{C0C0C0}\textbf{0}  & \cellcolor[HTML]{C0C0C0}\textbf{3} & \cellcolor[HTML]{C0C0C0}\textbf{2} & \cellcolor[HTML]{C0C0C0}\textbf{1} & \cellcolor[HTML]{C0C0C0}\textbf{0}  & \cellcolor[HTML]{C0C0C0}\textbf{1} & \cellcolor[HTML]{C0C0C0}\textbf{0} \\
\textit{Kerartoacanthoma in skin}            & \cellcolor[HTML]{C0C0C0}\textbf{2} & \cellcolor[HTML]{C0C0C0}\textbf{1}  & \cellcolor[HTML]{C0C0C0}\textbf{7} & \cellcolor[HTML]{C0C0C0}\textbf{0} & \cellcolor[HTML]{C0C0C0}\textbf{0} & \cellcolor[HTML]{C0C0C0}\textbf{1}  & \cellcolor[HTML]{C0C0C0}\textbf{0} & \cellcolor[HTML]{C0C0C0}\textbf{1}
\end{tabular}
\end{table}
In these data really near-to-linear shapes, such as $0,1,2,3$ tumor counts do not occur. I.e. a trend test in toxicology should be sensitive not only against near-to-linear shapes. Precisely for this reason an isotonic contrast test \cite{Peddada2006}, multiple contrast tests \cite{Rahman2017} a Williams-type contrast test \cite{Hothorn2010}, a sequential version of CA-test \cite{Rahman2017} were recommended as power-robust alternative. Not only the power argument is relevant. Whereas a contrast test considers \textit{dose} as a qualitative factor, the CA-test (and other regression-type tests) as a quantitative covariate. Sometimes, the consideration of 
\textit{control, low, medium, high} dose is appropriate, sometimes the administered dose levels $0,3,10,50$ mg/kg/wk. We will show both. And not only a certain sign of significance: p-value, or confidence limit for the appropriate chosen effect size, will be provided, but also the probable shape of the dose-response.
\textit{Therefore, the Tukey-Williams trend test is proposed here where dose is jointly considered quantitatively and qualitatively.}

For a mortality-adjusted analysis, the poly-3 modification should be available. Although p-values are recently widespread used, we should carefully consider the effect size, e.g. risk difference between two groups or odds ratio on the slope of a regression model. Furthermore, body or organ weight differences may influence the tumor incidences. Therefore, a related adjustment using the analysis of covariance should be available. A further special issue are downturn effects at high(er) dose(s) possibly caused by overdosing or limited bioavailability of the higher doses. To limit the false negative decision rate either pairwise comparisons against control or downturn-protected trend tests are highly recommended.\\

\section{A Tukey-Williams trend test}
Trend tests can be formulated for dose as either a qualitative factor or as a quantitative covariate. For the former a multiple contrast test can be used, for the latter the Armitage trend test of linear regression. Because a comparison against control is of interest, the Williams test \cite{Williams1971} is preferable as a special multiple contrast test. Because not only a linear regression is interested, the Tukey trend test \cite{Tukey1985} is used instead of the CA-trend test. Both tests represent a maximum test $T^{max}=max(T_1,T_2,...,T_\xi)$, whereby the tests $T_i$ represent studentized contrasts at the first, but at the second 3 regression tests for linear, ordinal and logarithmic dose scores. The global $T^{max}$ test, considering both multiple contrasts and 3 regression models, is jointly k-variate normal-distributed. However, the common correlation matrix is hard to calculate. The multiple marginal models approach (mmm) \cite{Pipper2012} provides the joint distribution of parameter estimates without assuming a certain multivariate distribution for the data. The variance-covariance matrix of parameter estimates is obtained using derivatives of the log likelihood function, i.e. without the explicit formulation of the correlation matrix in linear, generalized linear and generalized linear mixed effects models. \\
Because both a multiple contrast test and a regression test belong to the class of linear models, a specific maximum test for both the Williams contrasts and the Tukey regression models can be formulated $T^{max}_{\text{Tukey-Williams}}$ accordingly. It seems somewhat strange to answer one question about a trend alternative with $(3+k)$ individual tests. But first of all many possible forms of dose-response shapes are recognized with corresponding power: from linear to  plateau. Secondly, the test is not too conservative due to the high correlation between these models. Thirdly, the methodological break between quantitative and qualitative dose models can be overcome. Fourth, adjusted p-values and (compatible) simultaneous confidence intervals are available for this Union-Intersection-Test for individual inferences. This joint test principle is easily available in the CRAN package \verb|tukeytrend| \cite{Schaarschmidt2019}. The validity for the classes of lm, glm, glmm models allows very wide specifications, e.g. for adjusting covariates (like body weight in bioassays) or different effect sizes, like odds ratio and hazard ratio. On the other hand, this asymptotic procedure has coverage problems with small $n_i$.

\subsection{A Tukey-Williams trend test for crude proportions}
The Tukey-Williams trend test for proportions base on a the estimates of glm-model. The canonical link is the logit function with the odds ratio as effect size. Both standard Williams and Tukey test use the risk difference as effect size, which can be achieved by the identity link in the glm-model alternatively (under some data conditions). These asymptotic approaches reveal problems to keep the nominal coverage probability for small sample sizes, and the common used-NTP-design with $n_i=50,50,50,50$ is already \textit{small} under this perspective. Furthermore, ratio as effect size for multiple contrasts is rather unstable when considering $p_0=0$ which is a  relevant data outcome in long-term carcinogenicity bioassays. Both phenomena can be mitigated to some extend by the Add-2 data transformation \cite{Agresti2000}, simply adding one pseudo count to both tumor and no-tumor category in each group. Yes, for both p-value and confidence limit estimation a bias is introduced hereby, but the stabilizing effect of this rather simple modification is amazing \cite{Schaarschmidt2008}.\\
The realization of this new trend test by means of the CRAN package \verb|tukeytrend|  is quite simple, as demonstrated by the following example.
The 2-by-k-table data for incidence of squamous cell papilloma in male B6C3F1 mice administered 0, 0.0875, 0.175, 0.35, or 0.70 mM acrylamide \cite{Beland2013} are:

\begin{table}[ht]
\footnotesize
\centering
\begin{tabular}{rrrrrr}
  \hline
 & 1 & 2 & 3 & 4 & 5 \\ 
  \hline
Dose & 0 & 0.0875 & 0.175 & 0.35 & 0.70 \\ 
  No. papilloma & 0 & 2 & 2 & 6 & 6 \\ 
  No. animals at risk & 46 & 45 & 46 & 47 & 44 \\ 
   \hline
\end{tabular}
\end{table}

The simple R-code consists in the glm object for Add2-approximated proportions, the function \verb|tukeytrendfit| for Tukey and Williams test (the option \verb|ctype| stands for contrast type) and the function \verb|ghlt| (generalized linear hypothesis tests) from the package  \verb|multcomp| with the multiple marginal model approach (functions \verb|mmm, mlf|):
\scriptsize
\begin{verbatim}
library(tukeytrend)
library(multcomp)
swaOR <-glm(cbind(events+1,(n-events)+1)~dose, data=squam, family= binomial(link="logit")) 
tuOR <- tukeytrendfit(swaOR, dose="dose", 
				scaling=c("ari", "ord", "arilog", "treat"), ctype="Williams")
summary(glht(model=tuOR$mmm, linfct=tuOR$mlf, alternative="greater"))
\end{verbatim}
\normalsize

The adjusted p-values for the 7 models are shown in Table 2, whereas a regression model with ordinal dose scores 0,1,2,3,4 reveals the smallest p-value. The contribution of the Williams contrasts is weak, not surprising for an alternative pattern of $0,\zeta/3, \zeta/3, \zeta,\zeta$
\begin{table}[ht]
\centering
\scriptsize
\begin{tabular}{llrr}
  \hline
Model&Test & Test stats & p-value \\ 
  \hline
Covariate& arithmetic & 2.42 & 0.0168 \\ 
  & ordinal & 2.53 & 0.0126 \\ 
  & logarithmic & 2.53 & 0.0127 \\ 
Factor& C: 0- 0.70 & 1.96 & 0.0508 \\ 
  & C: 0-(0.35+0.70)/2 & 1.92 & 0.0545 \\ 
  & C:0- (0.175+0.35+0.70)/3 & 1.59 & 0.1050 \\ 
  & C: 0- (0.0875+0.175+0.35+0.70)/4 & 1.44 & 0.1378 \\ 
   \hline
\end{tabular}
\caption{Adjusted p-values} 
\label{tab:exa14}
\end{table}

Two further side notes: i) when $p_0=0$ and $n_i$ are small, the Add1 adjustment can be recommended to yield appropriate pseudo-score intervals \cite{Schaarschmidt2014}, ii) the choice of the effect size should be clear a priori, depending on the randomization, the design and the experimental goal. If this is not the case, this maximum test principle of highly correlated tests can be used to select one out of the three effect sizes:  difference of proportion, risk ratio or odds ratio that is most in line with the alternative \cite{Hothorn2020}.
 
\subsection{A Tukey-Williams trend test considering multiple tumors jointly}
The $T^{max}$ principle allows the consideration of any correlated tests, not only for between-dose inference, but for multiple correlated endpoints as well. Although multiple binary endpoints, such as multiple tumor site proportions are not too high correlated, are related multivariate 
$T^{max}_{\text{Tukey-Williams}}$ can be formulated using  the multiple marginal models approach (mmm).

As an example for multiple tumors data from the NTP bioassay on methyleugenol  (No. TR491) in female mice with tumors from 89 tumor sites were selected \cite{NTP2000}. The dataset contains the tumor counts together with the dose group (0, 37, 75, and 150 mg/kg) and the time of death \cite{Hothorn2015}. For demonstration purposes, six liver tumor sites were selected (abbreviated with \textit{t24: hepatoblastoma, t25: multiple hepatoblastoma, t27: multiple hepatocellular adenoma, t28: hepatocellular carcinoma, t29: multiple hepatocellular carcinoma, t30: hepatocholangiocarcinoma}

\begin{figure}[htbp]
	\centering
		\includegraphics[width=0.60\textwidth]{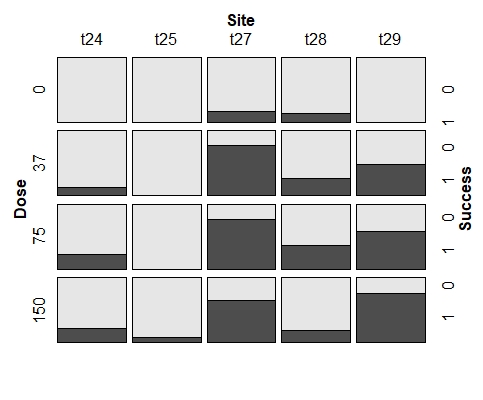}
		\caption{Mosaicplot multiple tumor incidence}
	\label{fig:mosaic5}
\end{figure}

Based on the six glm- and tukeyfit objects (tuxx) as in the above univariate case, the joint test is simply realized by the function \verb|combtt|:
\begin{verbatim}
tt5 <- combtt(tu24i, tu25i, tu27i, tu28i, tu29i)
summary(asglht(tt5, alternative="greater"))
\end{verbatim}

The multiplicity-adjusted p-values reveal no trend in either model or tumor site. If the analysis were only performed for the most significant tumor site and its best model alone, a strongly significant trend would emerge (last line of the table). This shows the extreme conservatism of multiple binary data in this approach. Therefore such a test should be used with care and not in routine.
\begin{table}[ht]
\centering
\scriptsize
\begin{tabular}{lllrr}
  \hline
Tumor site& Model&Test & Test stats & p-value \\ 
  \hline
t24& Covariate &arithmetic & 2.91 & 0.2888 \\ 
   & & ordinal & 3.15 & 0.2683 \\ 
  & & logarithmic & 3.15 & 0.2679 \\ 
  & Factor& C: 0- 150 & 2.58 & 0.3213 \\ 
  & & C: 0-(75+150)/2 & 2.58 & 0.3216 \\ 
  & & C: 0-(37+75+150)/3 & 2.35 & 0.3484 \\ 
  tu25 & Covariate& arithemic & 2.05 & 0.3926 \\ 
   & &ordinal & 1.82 & 0.4316 \\ 
   & & logarithmic & 1.82 & 0.4326 \\ 
  & Factor& C: 0- 150 & 1.52 & 0.4955 \\ 
  & & C: 0-(75+150)/2 & 0.72 & 0.7633 \\ 
  & & C: 0-(37+75+150)/3 & 0.49 & 0.8641 \\ 
  tu27 & Covariate & arithmetic & 3.62 & 0.2350 \\ 
  & & ordinal & 4.44 & 0.1934 \\ 
  & &logarithmic & 4.42 & 0.1950 \\ 
  & Factor&C: 0- 150 & 4.60 & 0.1878 \\ 
  & & C: 0-(75+150)/2 & 5.09 & 0.1696 \\ 
  & &  C: 0-(37+75+150)/3& 5.38 & 0.1608 \\ 
  tu28 & Covariate& arithmetic & 0.30 & 0.9416 \\ 
  & & ordinal & 0.80 & 0.7318 \\ 
  & & logarithmic & 0.80 & 0.7299 \\ 
  & Factor& C: 0- 150 & 0.52 & 0.8520 \\ 
  & & C: 0-(75+150)/2 & 1.43 & 0.5189 \\ 
  & & C: 0-(37+75+150)/3 & 1.43 & 0.5194 \\ 
  tu29 & Covariate &arithmetic & 6.41 & 0.1358 \\ 
  & & ordinal & 6.79 & 0.1280 \\ 
  & & logarithmic & 6.79 & 0.1288 \\ 
  &Factor & C: 0- 150 1 & 4.84 & 0.1780 \\ 
  & & C: 0-(75+150)/2 & 4.47 & 0.1935 \\ 
  & & C: 0-(37+75+150)/3& 4.22 & 0.2041 \\ \hline\hline
	tu29 alone & Covariate &logarithmic & 6.79 & $<0.0001$ \\ 
   \hline
\end{tabular}
\caption{Adjusted p-values} 
\label{tab:exa14}
\end{table}

\subsection{A Tukey-Williams test for poly-k estimates}
The poly-k test  represents a simple mortality-adjusted analysis where the cause-of-death information is not needed \cite{Bailer1988}. This adjustment performed well under several real data conditions and is robust to different tumor lethality patterns \cite{Moon2005}.
It accounts the dose-specific mortality by individual weights $w_{ij}=\left( t_{ij}/t_{max} \right)^k$ reflecting individual mortality pattern ($t_{ij}$ is time of death of animal $j$ in dose $i$).  The tuning parameter $k$ is data-dependent and in a recent NTP report they used in parallel $k=3$ and $k=6$. The weights result in adjusted sample sizes $n_{i}^{*}=\sum_{j=1}^{n_{i}}w_{ij}$ to be used instead of the randomized number of animals $n_{i}$. Therefore adjusted proportions $p_{i}^{*}=y_{i} /n_{i}^{*}$ are used instead of the crude tumor proportions $p_{i}=y_{i} /n_{i}$.  As an example the NTP study on the carcinogenic potential of methyleugenol for the incidence of skin fibromas is used \cite{NTP2000}.  By means of the CRAN package \verb|MCPAN| the $w_{ij}, n_{i}^{*},p_{i}^{*}$ can be easily estimated
  \begin{table}[ht]
	\footnotesize
    \caption{Crude and Poly-3-adjusted tumor rates in the skin fibroma example}
      \begin{center}
        \begin{tabular}{c|c|c|c|c}
        \hline
         dose & $0$ mg/kg & $37$ mg/kg & $75$ mg/kg & $150$ mg/kg \\
        \hline
        \hline
         Crude Rate &$1/50$&$9/50$&$8/50$&$5/50$\\ \hline
         Crude Percent &2.0\%&18.0\%&16.0\%&10.0\%\\
        \hline 
         Poly-3 adjusted Rate $y_{i} /n_{i}^{*}$ & $1/41.4$ & $9/40.3$ & $8/38.7$ & $5/32.7$ \\
         Poly-3 adjusted Percent $p_{i}^{*}$ & 2.4\% & 22.3\% & 20.7\% & 15.3\%\\
     \hline
         \end{tabular}
         \end{center}
         \label{Exa1adjusted}
    \end{table} 
The R code is for differences of proportions as effect size:
\footnotesize
\begin{verbatim}
fitpoly3 <- glm(tumour ~ dose, data=me, 
						family=binomial(link="identity"), weight=weightpoly3)
library(tukeytrend)
ttpoly3 <- tukeytrendfit(fitpoly3, dose="dose", 
				 	 scaling=c("ari", "ord", "arilog", "treat"), ctype="Williams")
glht(model=ttpoly3$mmm, linfct=ttpoly3$mlf, alternative="greater")
\end{verbatim}
\normalsize

\begin{table}[ht]
\footnotesize
\centering
\begin{tabular}{lllrrrrr}
  \hline
Poly-k&Model& Comparisons & Test stats & p-value  \\ 
  \hline
3& Covariate&arithmetic & 1.77 & 0.082 \\ 
  & ordinal& 2.50 & 0.015 \\ 
  & logarithmic & 2.48 & 0.017 \\ 
Factor& C: 0-150 & 1.91 & 0.062 \\ 
  & C: 0-(75+150)/2 &   3.03 & 0.0032 \\ 
  & C: 0-(37+75+150)/3 &  3.83 & 0.0002 \\ \hline
6&Factor	  & C: 0-(37+75+150)/3 &  3.97 & 0.00013 \\ 
   \hline
\end{tabular}
\caption{Poly-k trend test} 
\label{tab:exa14}
\end{table}
Notice, the mmm approach allows easily a maximum test over many correlated tests, not only over covariate and factor, over dose scores and contrasts, but also over several tuning parameter $k$.  In the last line of the table, the minimal p-value of such a 12-variate test is given, i.e. $k=6$ and the plateau alternative are the most likely. Using this example the robustness of the $T^{max}_{\text{Tukey-Williams}}$ against downturn effects at high doses is good to see.

\section{Biological gradient as key criterion}

That the presence of a significant dose-response dependency is used as a central criterion in the sense of Hill (biological gradient) in the evaluation of multiple glyphosate assays \cite{Greim2015} should be supported. But how was this actually implemented? Above, the analysis of a global pooled table (Table 20f) was discussed as inappropriate. The analysis of individual studies remains, here as an example Study 7 in Tab. 11 was selected. The separate evaluation of prematurely and terminally died animals is hardly comprehensible from a statistical point of view, but shows clearly different mortality patterns. This would have indicated a poly-k trend test, which unfortunately is not feasible due to missing single values. For hepatocellular adenoma in males the conclusion \cite{Greim2015} was \textit{'While not statistically significant using the Fishers exact test, the difference was statistically significant for total male rats using the Peto test for trend. However, there was no evidence of pre-neoplastic foci, no evidence of progression to adenocarcinomas,
and no dose-response.'} The 2-by 4 table is (corrected for $n_i\leq 52$):

\begin{table}[ht]
\footnotesize
\centering
\begin{tabular}{rrrrr}
  \hline
 & 1 & 2 & 3 & 4 \\ 
  \hline
Dose & 0 & 121 & 361 & 1214 \\ 
  No. hepatocellular adenoma & 0 & 2 & 0 & 6 \\ 
  No. animals at risk & 52 & 52 & 52 & 52 \\ 
   \hline
\end{tabular}
\caption{2-by 4 table hepatocellular adenoma} 
\end{table}

The  $T^{max}_{\text{Tukey-Williams}}$ provides  a p-value of $p^{arithmetic}=0.024$ (one-sided adjusted test for odds ratio in the Add-2 adjusted asymptotic approach). Although the dose-response pattern is not monotone, a significant trend exists and therefore a biological gradient in Hills sense should be claimed. Again, a poly-k adjusted analysis is highly recommended instead just crude proportions.
 
\section{Conclusion}
With the proposed $T^{max}_{\text{Tukey-Williams}}$ crude and mortality-adjusted tumor rates (using poly-k estimates) in long-term carcinogenicity bioassays can be evaluated appropriately. This test is sensitive against almost all dose-response shapes, from linear to specific non-monotone. \\
This test can be formulated for the three  effect sizes risk difference, risk ratio (not explicitly shown) and odds ratio. The test is asymptotic valid for a broad class of lm, glm, and glmm- models (latter not shown). To control the coverage probability, approximations, such as Add-2 , are recommended for the common small sample sizes, particularly when $p_0=0)$. Care is needed when considering low correlated tests, such as multiple tumors because here the conservatism may be unacceptable large.\\
Using the CRAN-packages \verb|tukeytrend, MCPAN, multcomp| various modifications of $T^{max}_{\text{Tukey-Williams}}$ are possible, demonstrated for several real data examples (the R-code is in the appendix).\\
Further extensions to consider multiple experiments and historical controls, as well as target concentrations vs. administered doses will be discussed next.

\footnotesize
\bibliographystyle{plain}

\scriptsize

\section{Appendix: R-Code including datasets}

\begin{verbatim}
library(xtable)
squam<-data.frame(
  dose = c(0,0.0875, 0.175, 0.35, 0.70),
  events = c(0, 2,2,6,6),
  n = c(46, 45, 46, 47,44))
tf<-t(squam)
rownames(tf)<-c("Dose", "No. papilloma", "No. animals at risk")
xtable(tf, digits=0)

library(tukeytrend)
swaOR <-glm(cbind(events+1,(n-events)+1)~dose, data=squam, family= binomial(link="logit"))# 
tuOR <- tukeytrendfit(swaOR, dose="dose", scaling=c("ari", "ord", "arilog", "treat"), ctype="Williams")
cOR <- summary(glht(model=tuOR$mmm, linfct=tuOR$mlf, alternative="greater"))
cOR$test$pvalues
library("ggplot2")
library(xtable)
COR<-fortify(cOR)[, c(1,5,6)]
colnames(COR)<-c("Test","Test stats", "p-value")
print(xtable(COR, digits=c(2,2,2,4), caption="Adjusted p-values", 
             label="tab:exa14"), include.rownames=FALSE)

############################# multiple tumors
library("reshape2")
data("miceF", package="SiTuR")
miceF$Dose[miceF$group==0]<-0
miceF$Dose[miceF$group==1]<-37
miceF$Dose[miceF$group==2]<-75
miceF$Dose[miceF$group==3]<-150
miceL <-miceF[, c(2, 27:28, 30:32, 93), ]

miceM <- melt(miceL, c("ID", "Dose"), variable.name="Site", value.name="Success")
library("plyr")
miceN <- ddply(miceM, .(Dose, Site), summarize, Successes=sum(Success))

library("vcd")
mosaic(Success~Dose+Site, data=miceM, zero_size=0, color=TRUE)

  library("multcomp")
glm24 <- glm(cbind(Successes + 1, 50 - (Successes+1)) ~ Dose, 
             subset(miceN, Site=="t24"), family=binomial())
glm25 <- glm(cbind(Successes + 1, 50 - (Successes+1)) ~ Dose, 
             subset(miceN, Site=="t25"), family=binomial())
glm27 <- glm(cbind(Successes + 1, 50 - (Successes+1)) ~ Dose, 
             subset(miceN, Site=="t27"), family=binomial())
glm28 <- glm(cbind(Successes + 1, 50 - (Successes+1)) ~ Dose, 
             subset(miceN, Site=="t28"), family=binomial())
glm29 <- glm(cbind(Successes + 1, 50 - (Successes+1)) ~ Dose, 
             subset(miceN, Site=="t29"), family=binomial())

tu24i <- tukeytrendfit(glm24, dose="Dose", scaling=c("ari", "ord", "arilog", "treat"), ctype="Williams")
tu25i <- tukeytrendfit(glm25, dose="Dose", scaling=c("ari", "ord", "arilog", "treat"), ctype="Williams")
tu27i <- tukeytrendfit(glm27, dose="Dose", scaling=c("ari", "ord", "arilog", "treat"), ctype="Williams")
tu28i <- tukeytrendfit(glm28, dose="Dose", scaling=c("ari", "ord", "arilog", "treat"), ctype="Williams")
tu29i <- tukeytrendfit(glm29, dose="Dose", scaling=c("ari", "ord", "arilog", "treat"), ctype="Williams")

tt5 <- combtt(tu24i, tu25i, tu27i, tu28i, tu29i)
TT5 <- summary(asglht(tt5, alternative="greater"))

library("ggplot2")
library(xtable)
T5<-fortify(TT5)[, c(1,5,6)]
colnames(T5)<-c("Test","Test stats", "p-value")
print(xtable(T5, digits=c(2,2,2,4), caption="Adjusted p-values", 
             label="tab:exa14"), include.rownames=FALSE)


tt24i <- tukeytrendfit(glm24, dose="Dose", scaling=c("ari", "ord", "arilog"))
tt25i <- tukeytrendfit(glm25, dose="Dose", scaling=c("ari", "ord", "arilog"))
tt27i <- tukeytrendfit(glm27, dose="Dose", scaling=c("ari", "ord", "arilog"))
tt28i <- tukeytrendfit(glm28, dose="Dose", scaling=c("ari", "ord", "arilog"))
tt29i <- tukeytrendfit(glm29, dose="Dose", scaling=c("ari", "ord", "arilog"))
tt30i <- tukeytrendfit(glm30, dose="Dose", scaling=c("ari", "ord", "arilog"))

tx<- combtt(tt24i, tt25i, tt27i, tt28i, tt29i)
stt10 <- summary(asglht(tx, alternative="greater"))

ta29i <- tukeytrendfit(glm29, dose="Dose", scaling="arilog") # alone
summary(glht(model=ta29i$mmm, linfct=ta29i$mlf, alternative="greater"))

############################ poly-k
library(MCPAN)
data("methyl", package="MCPAN")
data(methyl)
me <- methyl
# death: time of death, max(death)=length of study = 730 days
# poly-3 adjustment: (time of death/max(time))^k, k=3, for those animals without tumour at time of death!
# Compute the poly-3 (-k)- weights at the level of single animals
me$weightpoly3 <- 1
me$weightpoly6 <- 1
# Animals without tumour at time of death get corrected sample size
wt0 <- which(me$tumour == 0)
me$weightpoly3[wt0] <- (me$death[wt0]/max(me$death))^3
me$weightpoly6[wt0] <- (me$death[wt0]/max(me$death))^6
#me
# check n.adj of the weights per group (should be the same with poly3ci in MCPAN; see below: OK)
# Put the Poly-3-weighted GLM into tukeytrendfit:
# See help(methyl) for the original dose levels:
me$dosegroup <- me$group
levels(me$dosegroup) <- c("0", "37", "75", "150")
me$dose <- as.numeric(as.character(me$dosegroup))
# Notice, the  model use an identity link,
fitpoly3 <- glm(tumour ~ dose, data=me, family=binomial(link="identity"), weight=weightpoly3)
#summary(fitpoly3)
fitpoly3L <- glm(tumour ~ dose, data=me, family=binomial(link="logit"), weight=weightpoly3)
summary(fitpoly3L)
library(tukeytrend)
ttpoly3 <- tukeytrendfit(fitpoly3, dose="dose", scaling=c("ari", "ord", "arilog", "treat"), ctype="Williams")
compttpoly3 <- glht(model=ttpoly3$mmm, linfct=ttpoly3$mlf, alternative="greater")
ttpoly3L <- tukeytrendfit(fitpoly3L, dose="dose", scaling=c("ari", "ord", "arilog", "treat"), ctype="Williams")
compttpoly3L <- glht(model=ttpoly3L$mmm, linfct=ttpoly3L$mlf, alternative="greater")
library("ggplot2")
library(xtable)
polyL<-fortify(summary(compttpoly3L))
colnames(polyL)<-c("Comparisons","Test stats", "p-value")
print(xtable(polyL,  caption="Adjusted by multiple covariates", 
             label="tab:exa14"), include.rownames=FALSE)

polyI<-fortify(summary(compttpoly3))
colnames(polyI)<-c("Comparisons","Test stats", "p-value")
print(xtable(polyI,  caption="Adjusted by multiple covariates", 
             label="tab:exa14"), include.rownames=FALSE)
#########################poly 3 or 6 ###############################################

fitpoly6 <- glm(tumour ~ dose, data=me, family=binomial(link="identity"), weight=weightpoly6)
ttpoly6 <- tukeytrendfit(fitpoly6, dose="dose", scaling=c("ari", "ord", "arilog", "treat"), ctype="Williams")
compttpoly6 <- glht(model=ttpoly6$mmm, linfct=ttpoly6$mlf, alternative="greater")
tt36 <- combtt(ttpoly3,ttpoly6)
TT36 <- summary(asglht(tt36, alternative="greater"))
##########################################Hill criterion by Greim 2015
##########Study7 males hepatocell adenoma
library(xtable)
hep<-data.frame(
  dose = c(0,121, 361, 1214),
  events = c(0, 2,0,6),
  n = c(52, 52, 52, 52))
th<-t(hep)
rownames(th)<-c("Dose", "No. hepatocellular adenoma", "No. animals at risk")
xtable(th, digits=0)

library(tukeytrend)
hepOR <-glm(cbind(events+1,(n-events)+1)~dose, data=hep, family= binomial(link="logit"))# 
hOR <- tukeytrendfit(hepOR, dose="dose", scaling=c("ari", "ord", "arilog", "treat"), ctype="Williams")
HOR <- summary(glht(model=hOR$mmm, linfct=hOR$mlf, alternative="greater"))
HOR$test$pvalues
library("ggplot2")
library(xtable)
hor<-fortify(HOR)[, c(1,5,6)]
colnames(hor)<-c("Test","Test stats", "p-value")
print(xtable(hor, digits=c(2,2,2,4), caption="Adjusted p-values", 
             label="tab:exa44"), include.rownames=FALSE)

\end{verbatim}

\end{document}